\documentclass[proof]{WileyASNA-v1}

\usepackage{ulem}

\articletype{Article Type}%

\received{13 December 2019}
\revised{15 December 2019}
\accepted{17 December 2019}

\begin{document}

\title{
Accretion-induced collapse to third family compact stars as trigger for eccentric orbits of millisecond pulsars in binaries}

\author[1]{David Edwin Alvarez-Castillo*}
\author[2]{John Antoniadis}
\author[3,4]{Alexander Ayriyan}
\author[1,5,6]{David Blaschke}
\author[7]{Victor Danchev}
\author[3,4,8]{Hovik Grigorian}
\author[5]{Noshad Khosravi Largani}
\author[9,10]{Fridolin Weber}

\authormark{David Edwin Alvarez-Castillo, John Antoniadis, Alexander Ayriyan, David Blaschke, Victor Danchev, Hovik Grigorian, Noshad Khosravi Largani, and Fridolin Weber}

\address[1]{\orgdiv{Bogoliubov Laboratory for Theoretical Physics},
  \orgname{Joint Institute for Nuclear Research},
  \orgaddress{\state{Joliot-Curie street 6, 141980 Dubna,}
    \country{Russia}}}
\address[2]{
  \orgname{Max-Planck-Institut f\"ur Radioastronomie and Argelander Institut f\"ur Astronomie},
  \orgaddress{\state{Auf dem H\"ugel 69, 53121 Bonn,}
    \country{Germany}}}
\address[3]{\orgdiv{Laboratory for Information Technologies},
\orgname{Joint Institute for Nuclear Research},
\orgaddress{\state{Joliot-Curie street 6, 141980 Dubna,}
 \country{Russia}}}
\address[4]{\orgdiv{Computational Physics and IT Division},
\orgname{A.I. Alikhanyan National Science Laboratory},
\orgaddress{\state{Alikhanyan Brothers street 2, 0036 Yerevan},
\country{Armenia}}}
 \address[5]{\orgdiv{Institute of Theoretical Physics},
 \orgname{University of Wroclaw}, 
 \orgaddress{\state{Max Born place 9, 50-204 Wroclaw}, 
 \country{Poland}}}
\address[6]{
\orgname{National Research Nuclear University (MEPhI)},
\orgaddress{\state{Kashirskoe Shosse 31, 115409 Moscow,}
\country{Russia}}}
\address[7]{\orgdiv{Department of Physics},
\orgname{Sofia University St. Kliment Ohridski},
\orgaddress{\state{5 James Bourchier Blvd, 1164 Sofia}, 
\country{Bulgaria}}}
\address[8]{\orgdiv{Department of Physics},
\orgname{Yerevan State University}, 
\orgaddress{\state{Alek Manukyan street 1, 0025 Yerevan}, 
\country{Armenia}}}
\address[9]{\orgdiv{Department of Physics},
\orgname{San Diego State University}, 
\orgaddress{\state{5500 Campanile Drive, San Diego, CA
    92182}, \country{USA}}}
\address[10]{\orgdiv{Department of Physics},\orgname{University of
    California}, \orgaddress{\state{San Diego, La Jolla, CA 92093},
  \country{USA}}}  
\corres{*D. Alvarez-Castillo \email{sculkaputz@gmail.com}}

\abstract{
A numerical rotating neutron star solver is used to study the temporal evolution of accreting neutron stars
using a multi-polytrope model for the nuclear equation of state named ACB5.  
The solver is based on a quadrupole expansion of the metric, but confirms the results of previous works, revealing the possibility of an abrupt  transition of a neutron star from a purely hadronic branch to a third-family branch of stable hybrid stars, passing through an unstable intermediate branch. 
The accretion is described through a sequence of stationary rotating {stellar} configurations which lose angular
momentum through {magnetic} dipole emission while,
at the same time, gaining angular momentum through mass accretion.
The model has several free parameters which are inferred from observations.  
The mass accretion scenario is studied in dependence on the effectiveness of angular momentum transfer
which determines at which spin frequency the neutron star will become unstable against gravitational collapse
to the corresponding hybrid star on the stable third-family branch.
It is conceivable that the neutrino burst which accompanies the deconfinement transition may trigger a pulsar kick 
which results in the eccentric orbit.
A consequence of the present model is the prediction of a correlation between the spin frequency of 
the millisecond pulsar in the eccentric orbit and its mass at birth.  }

\keywords{neutron star, accretion, binary systems}

\maketitle

\section{INTRODUCTION}

\begin{figure}
\includegraphics[width=0.47\textwidth]{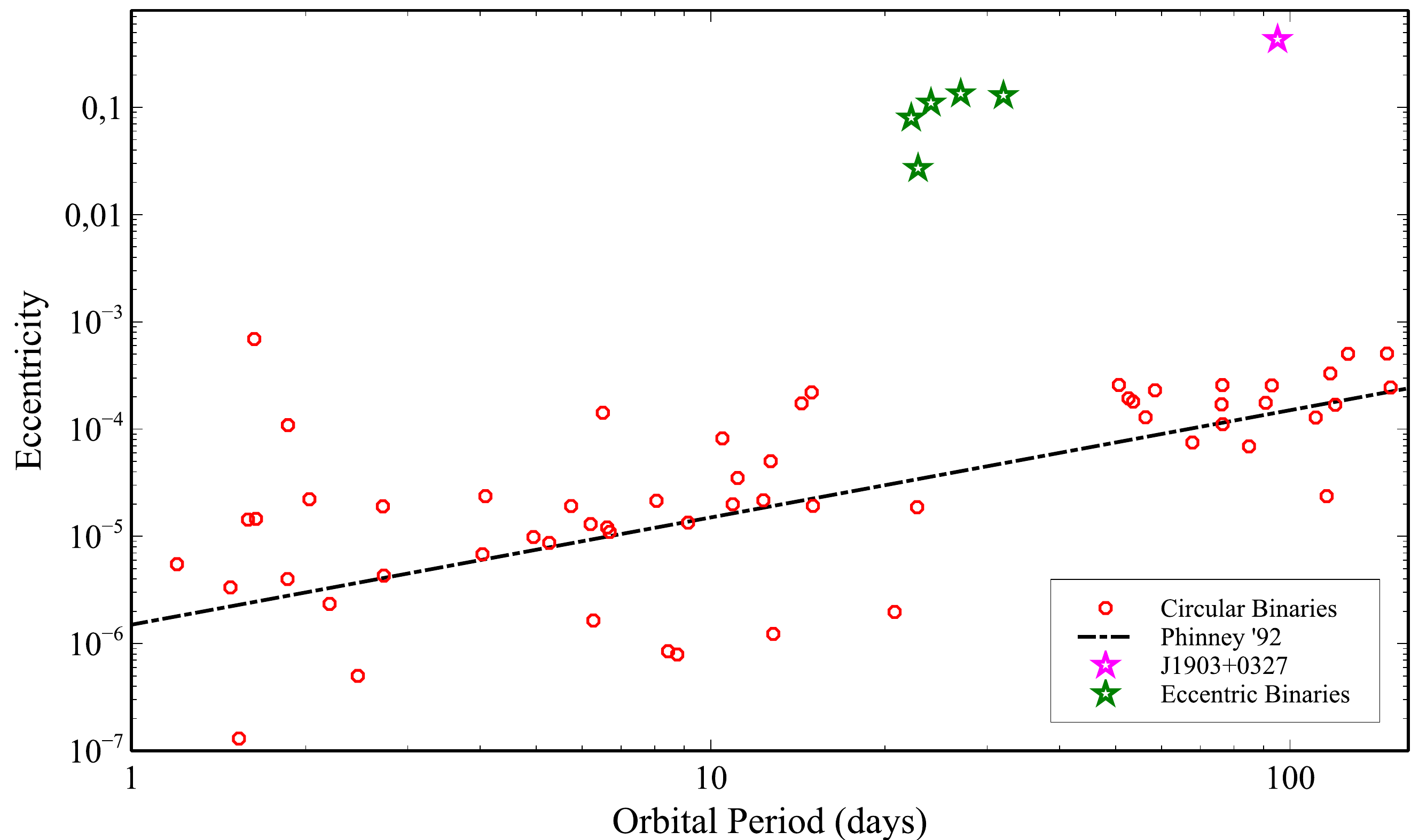}
\caption{Eccentricity vs. orbital period for millisecond pulsars in binaries with white dwarf companions, see~\citep{Antoniadis:2014,Stovall:2018rvy}.
}
\label{fig:eMSP}
\end{figure}

One of the most intriguing phenomena in compact star physics is the appearance of eccentric orbits in binaries of a millisecond pulsar (MSP) with a low-mass white dwarf in a rather narrow region of orbital periods, see Fig.~\ref{fig:eMSP}.
In \citep{Antoniadis:2014} it has been demonstrated that the observed eccentricities of $e\simeq 0.01 - 0.15$ can be generated  from the dynamical interaction of the binary with a circumbinary disc for periods between 15 and 50 days.
\cite{Freire:2013xma}, however, proposed that these binary MSPs may form from the rotationally delayed accretion-induced collapse of a massive white dwarf. 
In \cite{Antoniadis:2016bnj} it has been shown that the companion star to one of these MSPs is indeed a low-mass white dwarf.
In a recent work \citep{Jiang:2015gpa} an accretion induced phase transition to a strange quark star has been suggested as an origin for the eccentricity of the orbit in the period range. 

In this contribution we would like to explore a scenario of accretion onto a pulsar that may undergo a phase transition to a hybrid star with a quark matter core situated on a third family branch which thus involves a decent spin-up and structural reconfiguration of the compact star interior that may result in energy release by violent neutrino emission.
The latter may result in the neutrino rocket mechanism that produces a compact star kick (see, e.g., \citep{Peng:2003hr,Berdermann:2005yn,Sagert:2007as}) as the origin for the eccentricity just in that window of orbital periods where mass transfer from a circumbinary disc is most effective.

In order to develop the scenario, we shall first consider a model for disc accretion that results in spin-up and mass increase of the compact star. We then suggest an equation of state (EoS) with a strong phase transition for the compact star matter fulfilling the criterion for the formation of a third-family branch of hybrid stars \citep{Gerlach:1968zz}, separated from the one for ordinary neutron stars by a sequence of unstable configurations \citep{Alford:2013aca}. 
With such a setting one can suggest a prompt catastrophic rearrangement of the stellar structure as the origin for a compact star kick and thus an eccentric orbit.
The EoS shall be chosen such that the modern constraints for the minimal maximum mass of 
$\sim 2~M_\odot$ for PSR J0740+6620 \citep{Cromartie:2019kug} and the new NICER mass-radius constraints for PSR J0030+0451 will be fulfilled: $M=1.44^{+0.15}_{-0.14}~M_\odot$ and $R=13.02^{1.24}_{1.06}$ km for three oval spots \citep{Miller:2019cac} as well as $M=1.34^{+0.15}_{-0.16}~M_\odot$ and $R=12.71^{1.14}_{1.19}$ km for one small hot spot and a azimuthally-extended narrow crescent \citep{Riley:2019yda}.
Moreover, the EoS should describe a pulsar at $1.4~M_\odot$ with a tidal deformability obeying the LIGO constraint 
$\Lambda_{1.4}=190^{+390}_{120}$ \citep{Abbott:2018exr}.
Since the lowest well-measured mass for the eccentric MSPs in Fig.~\ref{fig:eMSP} is $1.353^{+0.014}_{0.017}~M_\odot$
for PSR J2234+0611 \citep{Stovall:2018rvy}, the candidate EoS for the scenario should have in the static case an onset mass 
for the strong phase transition at or below this value.  

\section{METHODOLOGY}
\label{sec:method}
The results presented in this paper are obtained with the
authors' own numerical integrator of Einstein's field equation for
rotating compact stars based on the Hartle and Thorne
methodology~\citep{Hartle:1968si}, which consists of the
development of a perturbation solution based on the Schwarzschild metric.  
The perturbed metric is expanded through second order in the star's rotational velocity, $\Omega$, and is given by
\begin{eqnarray}\label{metric}
ds^2 & \!\! = \!\! & - e^{\nu}\Bigl[ 1 \! + \! 2( h_0 + h_2 \mathit{P}_{2} ) \Bigr]dt^2 \! + \!
e^{\lambda}\left[ 1 + \frac{2( m_0 + m_2 \mathit{P}_{2})}{r - 2 M}
  \right] dr^2 \nonumber  \\ & \!\! + \!\! & r^2 \Bigl[1 \! + \!  2(v_2 - h_2)\textit{P}_2\Bigr] \Bigl[d\theta^2 \! + \! 
  \sin^2( d\phi - \omega dt)^2 \Bigr] \! + \!  O(\Omega^3), \nonumber \\
\end{eqnarray}
where the functions $M(r)$ and $\lambda(r)$ are solutions of the
Tolman-Oppenheimer-Volkoff (TOV) equation 
($e^\lambda = ( 1 - 2M/r)^{-1}$), and $h_0(r)$, $m_0(r)$, $h_2(r), m_2(r)$, and $v_2(r)$
denote monopole and quadrupole perturbation functions.  
The stellar deformation is described by the Legendre polynomial 
\begin{equation} \textit{P}_2 = \textit{P}_2(\cos{\theta}) = (3 \cos^2{\theta} - 1)/2 \, , \end{equation} and
$\omega(r)$ is the angular (frame-dragging) frequency relative to a
static observer at spacelike infinity.  
The energy-momentum tensor is characterized by two more perturbation functions, $p_0^*(r)$ and
$p_2^*(r)$, once again contributing to radial and $\theta$-dependent
perturbations (through $\mathit{P}_2$).  
The equations of all the perturbations can be readily derived from Einstein's field
equations~\citep{Hartle:1968si}.  
The perturbation equations can be integrated by starting from the TOV solution by choosing a guess value
for the angular velocity and fixing it through multiple integrations with the shooting method. 
The angular velocity for an inertial local
observer $\bar{\omega}$ is obtained in this way.  It is related to the
angular frequency $\omega$ in \eqref{metric} through $\bar{\omega} =
\Omega - \omega$, where $\Omega = u^{\phi}/u^t$ is the star's angular
velocity (defined by the 4-velocity components of a co-rotating observer).  
The asymptotics of the perturbation functions allow one to
find them unambiguously for a given angular velocity
$\Omega$ or angular momentum boundary value $J$. 

The code is conceptually divided into a TOV integrator and a function
which takes the TOV input and integrates all the perturbations
radially outward with the corresponding boundary value
settings~\citep{Hartle:1968si}.  A $4^{\rm th}$ order Runge-Kutta
method has been used due to its relative simplicity and accuracy.
Several versions of the code were developed, one of which
allowing us to determine the  properies of stars with pre-defined baryon numbers
(listed further down in the paper).
In addition to the standard procedure of Hartle and Thorne, care has
been taken to compute the proper mass shedding limit by taking into account the
dragging of inertial frames as described in~\citep{Weber:1999qn}.

The model for matter accretion from a neutron star's
disc is based on~\citep{Bejger:2011ir}, considering the baryon mass
$M_b$ and angular momentum $J$ to be the variables of
such an evolution.  A single star's evolution is seen as a set of
stationary rotating configurations of different $M_b$ and $J$ values.  
Angular momentum can be gained from infalling
particles and lost through the interaction between the star's magnetic
field and the accretion disk.  The relevant evolution equation is
\begin{equation}
\label{Jevolution}
\frac{dJ}{dM_b} = l_{\mathrm{tot}} = \kappa l(r_0) - l_m ,
\end{equation}
where $l(r_0)$ is the specific angular momentum of the infalling
particle at the innermost radius of the  disc, $r_0$,
while $l_m$ gives the magnetic torque divided by the accretion rate.
The quantity $\kappa$ describes the amount of angular
momentum of the infalling particles which the star can actually
receive. The  value of $\kappa$ is thus in the range of $[0,1]$.
The magnetic torque is given in terms of the dipole moment of the star, $\mu = BR^3$ as
\begin{equation}
l_m = \frac{\mu^2}{9 r_0^3 \dot{M}_{b}} \left( 3 -
2\sqrt{\frac{r_c^3}{r_0^3}} \right).
\end{equation}
The inner disc radius is determined through a transcendental equation
relating the magnetospheric radius $r_m$, the corotation radius $r_c$
and the relativistic marginally stable orbit $r_{ms}$ as follows
\begin{equation}
\frac{1}{\Omega r}\frac{d l}{dr} = \frac{1}{2}f_{ms}(r_0) = \left(
\frac{r_m}{r_0} \right)^{7/2} \left( \sqrt{\frac{r_c^3}{r_0^3}} - 1
\right),
\end{equation}
where the corresponding radii are given by
\begin{eqnarray}
\label{eq:r_m}
r_m &=& \frac{\mu^\frac{4}{7}}{(GM)^{\frac{1}{7}}\dot{M}_b^{\frac{2}{7}}} \, , \\
r_c &=& \left( \frac{GM}{\Omega^2} \right)^{1/3} \, , \\
0  &=& f_{ms}(r_{ms})  .
\end{eqnarray}
Evolution through accretion is modelled by numerical integration of
\eqref{Jevolution} which makes use of the boundary $J$ for a given
$M_b$ integration of neutron star models to generate a sequence based
on initial baryon mass and magnetic field values.  The results further
down were obtained by preparing an array of $J$ values and integrating
\eqref{Jevolution} with the corresponding $M_{b0}$ values and  a given value
of the angular momentum transfer coefficient $\kappa$.  Last but
not least, a decrease of the neutron star magnetic field has been
considered through a simple phenomenological model in which the
accreted mass $M_{acc}$ "buries" the initial magnetic field $B_i$,
described by
\begin{equation}
B(M_{acc}) = \frac{B_i}{1 + M_{acc}/m_B}.
\end{equation}
This model was suggested  by~\citep{Shibazaki:1989nature}. 
It has a free parameter $m_B$ which controls the speed at which the magnetic
field decreases.  Values of $m_B$ based on observations are in the
range $10^{-5}$ -- $10^{-4} M_{\odot}$.

A final estimate on the time during which accretion takes place
depends on the value of $\dot{M}_b$.  It should be noted that the
overall evolution is degenerate since the ratio $\mu^2/\dot{M}_b$
determines the RHS of Eq. (\ref{eq:r_m}).  
Changing the values of $\mu$ and $\dot{M}_b$ simultaneously can lead to the same
track if properly scaled, but the time in which the overall evolution
takes place depends only on $\dot{M}_b$.  
For that reason, several different values of  the initial magnetic field $B_i$ and
$\dot{M}_b$ have been considered in the listed results to show the dependence on these quantities.  
All of them, however, are of the order of $\dot{M}_b \cong 10^{-9} M_{\odot}$, $ B_i \cong 10^{8}$ G. 

\section{EQUATION OF STATE}
For the present study we will exploit an equation of state in the form of a multi-polytrope parametrisation as it
can be found, e.g., in \citep{Read:2008iy,Hebeler:2013nza,Alvarez-Castillo:2017qki}. 
It consists of piecewise polytrope regions in density that are joined thermodynamically consistently. 
For the lowest density parts that correspond to the outer NS crust ($n < 0.5 n_0$) we implement the BPS EoS
(\citep{Baym:1971pw}), to be followed by an intermediate phase of homogeneous matter in $\beta-$ equilibrium 
($0.5 n_0 < n < 1.1 n_0$). Above $1.1 n_0$, we introduce four polytropic segments described by:
\begin{eqnarray}
P(n)=\kappa_{i}(n/n_{0})^{\Gamma_{i}} \, ,
\label{polytrope}
\end{eqnarray}
where $n_{i}<n<n_{i+1},\ i=1\dots4$  define the regions with constant $\kappa_{i}$ and polytropic index $\Gamma_{i}$.  Since $P(n) =  n^{2}(d(\varepsilon(n)/n))/(dn)$,
the remaining thermodynamical variables can be easily determined \citep{Zdunik:2005kh}:
\begin{eqnarray}
\varepsilon(n)/n & = &
\frac{1}{n_{0}^{\Gamma_i}}\int dn\,\kappa
n^{\Gamma_i-2}
=
\frac{1}{n_{0}^{\Gamma_i}}\frac{\kappa\,n^{\Gamma_i-1}}{\Gamma_i-1}+C,\\ \mu(n)
& = &
\frac{P(n)+\varepsilon(n)}{n}=\frac{1}{n_{0}^{\Gamma_i}}\frac{\kappa\,
  \Gamma_i}{\Gamma_i-1}n^{\Gamma_i-1}+m_{0} \label{5} \, , 
\end{eqnarray}
where $C$ is fixed by the condition$\varepsilon(n\to0)=m_{0}\,n$. 
The above equations can be recast into the following forms:
\begin{eqnarray}
n(\mu) & = &
\left[n_{0}^{\Gamma_i}(\mu-m_{0})\frac{\Gamma_i-1}{\kappa\Gamma_i}\right]^{1/(\Gamma_i-1)} \!\!\!\!\! \!\!\!\!\! \!\!\!\!\! \!\!   ,
\end{eqnarray}
\begin{eqnarray}
P(\mu) & = & \kappa\left[n_{0}^{\Gamma_i}(\mu-m_{0})\frac{\Gamma_i-1}
  {\kappa\Gamma_i}\right]^{\Gamma_i/(\Gamma_i-1)} ~\label{P-mu} \!\!\!\!\! \!\!\!\!\! \!\!\!\!\! \!\!\!\!\!  , 
\label{eq:P-mu}  
\end{eqnarray}
which are suitable for performing a Maxwell construction of a first-order phase transition between the
hadron and quark phases.
Each of the polytropic regions has physical meaning. The first one is defined as the result of a fit to the stiffest EoS version provided in \citep{Hebeler:2013nza}. 
We are interested in an EoS with a strong first order phase transition that would produce a large jump in energy density 
$\Delta \varepsilon$ at the critical pressure $P_{c}$ in order to fulfill the Seidov criterion \citep{Seidov:1971}
for a gravitational instability: $\Delta \varepsilon > (\varepsilon_c + 3P_c)/2$. 
Therefore we define the second polytrope as a region of constant pressure $P_{c}=\kappa_{2}$ ($\Gamma_{2}=0$), 
which can result from a Maxwell construction using (\ref{eq:P-mu}) for the neighboring regions. 
The remaining polytropes, in the regions $i=3, 4$, at the highest densities beyond the first order phase transition shall correspond to stiff quark matter. 
It shall be sufficiently stiff to result in stable hybrid star configurations with a maximum mass of at least  $2~M_\odot$, 
thus fulfilling the constraint from the lower bound of the $2\sigma$ region of PSR J0740+6620 \citep{Cromartie:2019kug}, and similar constraints stemming from PSR J0348+0432  \citep{Antoniadis:2013pzd} and PSR J1614-2230 
\citep{Demorest:2010bx}. 

\begin{table}[!htb]\centering 
\caption{ EOS for the ACB5~\citep{Paschalidis:2017qmb}.
The functional form of the polytrope is defined by Eq.~(\ref{polytrope}) in the main text.
The first polytrope ($i=1$) is fitted to the nuclear EoS at supersaturation densities, the second one ($i=2$) 
describes a first-order phase transition with constant pressure $P_{c}=\kappa_2$ and $\Gamma_2=0$ in the 
$n_{2} < n < n_{3}$ region. The polytropes that lie in regions 3 and 4, i.e., above the phase transition represent 
high-density matter, e.g., quark matter.  Maximum masses $M_{{\rm max}}$ on the hadronic
and hybrid branches correspond to region 1 and 4. 
The minimal mass $M_{{\rm min}}$ on the hybrid branch lies in region 3 and marks the onset of the third family solution.} {%
\begin{tabular}{c|cccc|c}
\hline 
  & $\Gamma_{i}$  & $\kappa_{i}$  & $n_{i}$  & $m_{0,i}$  & $M_{{\rm max/min}}$\tabularnewline
 i  &  & {[}MeV/fm$^{3}${]}  & {[}1/fm$^{3}${]}  & [MeV]  & $[M_{\odot}${]}\tabularnewline
\hline 
 1  & 4.777  & 2.1986  & 0.1650  & 939.56  & 1.40 \tabularnewline
 2  & 0.0  & 33.969  & 0.2838  & 939.56  & -- \tabularnewline
 3  & 4.000  & 0.4373  & 0.4750  & 995.03  & 1.39 \tabularnewline
 4  & 2.800  & 2.7919  & 0.7500  & 932.48  & 2.00 \tabularnewline
\hline 
\hline 
\end{tabular}
\label{param-45} 
} 
\end{table}

As it has been noted in \citep{Alvarez-Castillo:2017qki}, it is rather challenging to find a parametrisation with only one polytrope for the high-density region that would obey thermodynamic consistency and at the same time provide a high maximum mass without violating the causality constraint on the squared speed of sound $c_s^2 = dP/d\varepsilon < 1$.
The multi-polytrope parametrisation has also proven successful for analyses of the compactness constraint that follows from the recent measurements of tidal deformability \citep{Abbott:2018exr} (see also \citep{De:2018uhw,Zhao:2018nyf}) 
of compact stars in the mass range around $1.36~M_\odot$ by analysing the gravitational wave signal from the inspiral phase of the binary merger GW170817 \citep{Annala:2017llu,Paschalidis:2017qmb}   and by placing constraints on the 
EoS \citep{Miller:2019nzo}.

The four-polytrope model which was introduced in \citep{Paschalidis:2017qmb} and denoted there as "ACB5" fulfills the present constraints from multimessenger astronomy on maximum mass and compactness of compact stars (given at the end of the Introduction) and has recently also been studied concerning its behaviour under fast rotation 
\citep{Bozzola:2019tit,Blaschke:2019tbh}. 
In Table \ref{param-45} we provide the parameters of this EoS.
 
\section{RESULTS, DISCUSSION AND PERSPECTIVES}
We have utilized our rotating neutron star solver in order to obtain solutions for the axially symmetric solutions
of the Einstein equations with the EoS ACB5. 
On this basis we studied the spin and mass evolution of the compact star within the accretion scenario defined in 
Section \ref{sec:method} in dependence on the effectiveness of angular momentum transfer, as described also  in 
\citep{Bejger:2016emu} for a similar case.

\begin{figure}
\includegraphics[width=0.54\textwidth]{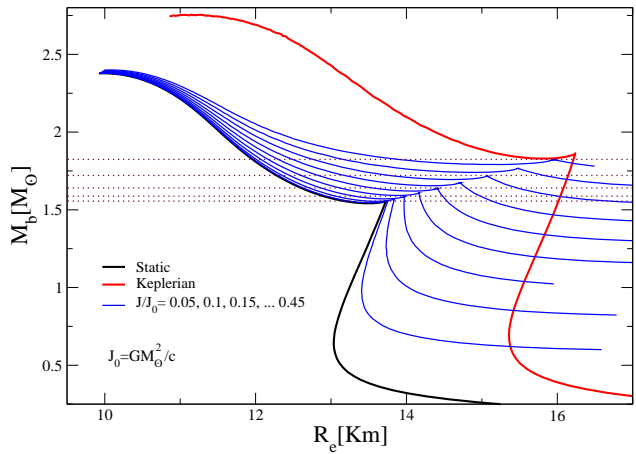}
\caption{Baryon mass versus equatorial radius for the ACB5 EoS.
Static configurations are black solid lines, the Keplerian limit is shown as red solid line.
Blue solid curves are at constant angular momentum $J$.
The maroon horizontal dotted lines of constant baryon mass can guide the eye to
follow from an endpoint of stability on the hadronic branch to a point on the hybrid 
star branch which would be reached by a collapse under simultaneous conservation 
of angular momentum $J$ and baryon mass $M_b$. 
For the gravitational mass of these hybrid star configurations, see the light green
curve in Fig.~\ref{fig:M-R}.
}
\label{fig:Mb-R}
\end{figure}

In Fig.~\ref{fig:Mb-R} we show the baryon mass versus equatorial radius for given angular momenta 
$J=0.05, 0.1, 0.15, \dots , 0.45~J_0$, with $J_0=GM_\odot^2/c$ (blue solid lines), together with the static case 
(black solid line) and the Keplerian limit (red solid line).  
In order to guide the eye we have shown also a few lines of constant baryon mass which allow to estimate the 
path of a collapsing star from the edge of the hadronic branch to the corresponding point on the stable hybrid star branch 
(third family) under simultaneous conservation of angular momentum and baryon mass.
We want to note that for the present type of EoS such a catastrophic rearrangement of the stellar interior is possible,
in principle, at the free-fall time scale.
In previous works on quark deconfinement in accreting low-mass binaries \citep{Glendenning:1997fy}
this was not the case and the rearrangement 
would have taken a secular time scale of about $10^6 - 10^8$ yr in order to get rid of the angular momentum difference
by, e.g., electromagnetic dipole radiation.
However, such a direct transition scenario would also entail that the orbits would re-circularize quite fastly (on a scale of
$\sim 10^3$ yrs) due to mass-accretion and/or tidal forces. 
The collapse to a hybrid star as an origin for the eccentric MSP orbit in the prompt collapse scenario should therefore have occurred rather recently.

\begin{figure}
\includegraphics[width=0.53\textwidth]{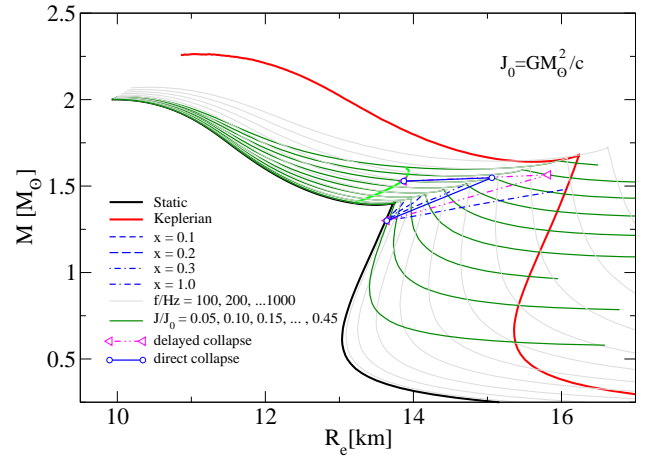}
\caption{Gravitational mass versus equatorial radius for the ACB5 EoS.
Static configurations are black solid lines, the Keplerian limit is shown as red solid line.
Green curves are at constant angular momentum $J$ and grey lines are for constant 
rotation frequency $f=1/P$, where $P$ is the period. 
The light green curve connects those points on the 
hybrid star branches which can be reached by a collapse from the maximum of stability 
on the hadronic branch under conservation of angular momentum $J$ and baryon mass 
$M_b$, see Fig.~\ref{fig:Mb-R}.
The blue lines with different styles indicate evolution trajectories that stars take in this diagram 
according to the accretion model describes in the text, for different efficiencies $x$ of 
angular momentum transfer. The solid blue line corresponds to direct collapse while the magenta 
dash-double-dotted line shows the path for delayed collapse.}
\label{fig:M-R}
\end{figure}

Alternatively, in order to circumvent the problem with re-circularization in the prompt collapse scenario, one can discuss a delayed collapse scenario on the basis of the calculation presented in Fig.~\ref{fig:M-R}.
If the binary star gains mass by accretion but fails to reach the onset mass for the transition when the accretion and spin-up process stops, it could then undergo a spin-down evolution and thus reach the instability line for transition to a hybrid star configuration by loss of angular momentum.
There are several realisations of this process, depending on the period of the binaries: a)
short period binaries which would always reach the maximum mass during accretion, resulting in circular orbits, b) long
period binaries that would collapse only after spin-down, producing eccentric orbits, and c) very long period binaries
that would never undergo the transition.

In Fig.~\ref{fig:M-R} we show the gravitational mass versus equatorial radius for the same ACB5 EoS and for the same grid of fixed angular momenta (green solid lines). 
The light grey lines are curves of fixed spin frequency $f=100, 200, \dots , 1000$ Hz.
The blue lines with different styles are the trajectories that a pulsar takes under 
mass accretion during the low-mass X-ray binary phase for given effectiveness $x$ of the angular momentum transfer in this process.  Once the trajectory reaches the line that joins the endpoints of stability of hadronic star configurations, the collapse sets in which transforms the hadronic star into a hybrid star, with a birth mass on the light green line.
We note that for a very effective angular momentum transfer (e.g., $x=1$) the star would never reach to the line for the onset 
of collapse. It would instead continue the spin-up process until it goes over to the so-called propeller regime.

\begin{figure}
\includegraphics[width=0.53\textwidth]{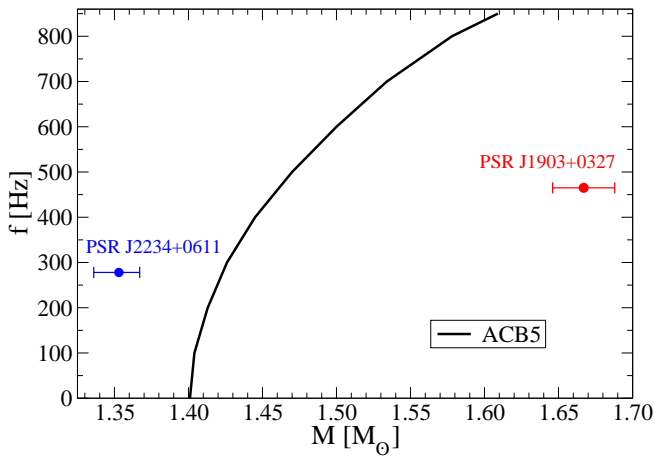}
\caption{Spin frequency versus gravitational mass of the hybrid star at its birth in the accretion 
induced collapse from a neutron  star. 
The black solid line shows the result of our model for the ACB5 EoS.
Of the few known millisecond pulsars in eccentric orbits (see Fig.~\ref{fig:eMSP}) we show those 
for which mass and spin are precisely measured: PSR J2234+0611 \citep{Stovall:2018rvy} and 
PSR J1903+0327 \citep{Freire:2010tf}. 
They confirm the trend of the present model: larger mass entails higher spin frequency.}
\label{fig:f-M}
\end{figure}

In Fig.~\ref{fig:f-M} we show as the black solid line the final result of this work, the correlation between frequency and mass of the spun-up MSP that is kicked into an eccentric orbit by a strong phase transition, here described by the model EoS ACB5.
We note that this result is strongly EoS dependent and could thus serve as a diagnostic tool for determining the state of 
matter in compact stars. 
In particular, the lower limit of the curve depends on the onset mass of the phase transition for the static configuration as a solution of the Tolman-Oppenheimer-Volkoff equations. Here, a lower onset mass for the third-family branch (and thus the so-called mass twins) could be provided, e.g., by the hybrid star EoS on the basis of the  generalized nonlocal NJL model EoS for color superconducting developed in \citep{Alvarez-Castillo:2018pve} guided by the density-functional based quark matter EoS for the relativistic string-flip model of \citep{Kaltenborn:2017hus}.
For such an EoS it would be no problem to match the data point for PSR J2234+0611  \citep{Stovall:2018rvy} also shown in that figure.

It should be noted that the high-mass pulsar PSR J1903+0327 has a highly debated origin \citep{Freire:2010tf} and should probably not be considered to belong to the population of the other eccentric MSPs shown in Fig.~\ref{fig:eMSP} since its companion is not a low-mass white dwarf but rather a main sequence star and its orbital period of 95.2 days is beyond the 
range of 15 - 50 days for the other eccentric MSPs.
In order to put our scenario for the origin of the eccentric MSPs to the test, we look forward to precise mass measurements, 
in particular for highly spun-up pulsars like PSR J0955-6150 with a spin period of 2.0 ms \citep{Octau:2018rdq}. 

For the present scenario it is not immediately obvious how more massive eMSPs could be explained like 
PSR J1946+3417 \citep{Barr:2016vxv} with a precisely determined high mass of $1.828(22)~M_\odot$.
It is, however, also possible to develop the present scenario further so that another major structural rearrangement contributing to a pulsar kick may take place, even leading from a third to a fourth family, see \citep{Alford:2017qgh,Li:2019fqe}.


\subsection*{Acknowledgements}
We acknowledge  the Russian Science Foundation for supporting the part of the work on phenomenology 
of compact stars with strong first order phase transition under grant number 17-12-01427. 
The work of F.W. was supported by the National Science Foundation (USA) under Grant PHY-1714068.

\end{document}